\documentclass[onecolumn,  preprint, 12pt, letterpaper, showpacs]{revtex4}
\usepackage{amstext}
\usepackage{amsmath}
\usepackage{amssymb}
\usepackage{graphicx}

\makeatletter
\def\@dotsep{4.5}
\makeatother

\newcommand{\ud}{\mathrm{d}}
\newcommand{\vect}[1]{\boldsymbol{#1}}
\newcommand{\eq}[1]{Eq.~\eqref{#1}}
\newcommand{\fig}[1]{Fig.~\ref{#1}}
\newcommand{\stn}[1]{Sec.~\ref{#1}}
\newcommand{\be}{\begin{equation}}
\newcommand{\ee}{\end{equation}}
\newcommand{\ba}{\begin{align}}
\newcommand{\ea}{\end{align}}
\newcommand{\ti}[1]{\text{#1}}
\newcommand{\mc}[1]{\mathcal{#1}}

\newcommand{\cb}{CBPQT$^{4+}$}

\begin{document}
\title{AFM pulling and the folding of donor-acceptor oligorotaxanes: phenomenology and interpretation}
\author{Ignacio Franco}
\author{George C. Schatz}
\author{Mark A. Ratner}
\affiliation{Department of Chemistry, Northwestern University, Evanston, Illinois 60208-3113}

\date{\today}

\begin{abstract}
The thermodynamic driving force in  the self-assembly of the secondary structure of a class of donor-acceptor oligorotaxanes is elucidated by means of molecular dynamics simulations of equilibrium isometric single-molecule force spectroscopy  AFM experiments. The oligorotaxanes consist of cyclobis(paraquat-\emph{p}-phenylene) rings threaded onto an oligomer of 1,5-dioxynaphthalenes linked by polyethers. The simulations are performed  in a high dielectric medium using MM3 as the force field.   The resulting  force vs. extension isotherms show a mechanically unstable region in which the molecule unfolds and, for selected extensions,  blinks in the force measurements between a high-force and a low-force regime. From the force vs. extension data the molecular potential of mean force is  reconstructed using the weighted histogram analysis method and decomposed into energetic and entropic contributions. The simulations  indicate that the folding  of the oligorotaxanes is energetically favored but entropically penalized, with the energetic contributions overcoming the entropy penalty and effectively driving the self-assembly.  In addition, an analogy between the single-molecule folding/unfolding events driven by the AFM tip and the thermodynamic theory of first-order phase transitions is discussed and general conditions, on the molecule and the cantilever, for the emergence of mechanical instabilities and blinks in the force measurements  in equilibrium isometric pulling experiments are presented. In particular, it is shown that the mechanical stability properties observed during the extension are intimately related to the fluctuations in the force measurements. 
\end{abstract}


\maketitle

\section{Introduction}

Unfolding and unbinding events in individual macromolecules  can be  
computationally studied by means of ``pulling" molecular dynamics (MD)  
simulations~\cite{smd1, grubmuller99, smd2}.  These calculations simulate  recent single-molecule 
experiments in which a molecule (typically a protein or DNA) attached to a surface is  
mechanically unfolded by pulling it with  optical tweezers or an atomic force microscope (AFM)  
tip attached to a cantilever~\cite{evans2001, Liphardt01, bustamanterev05,   ritort06}.  Such experiments show stress maxima that have 
been linked to the breaking of hydrogen bonds, providing insight into the 
secondary and tertiary structure of the macromolecules, and the dynamical processes involved in muscle contraction, protein folding, transcription, shape memory materials and many others.

\begin{figure}[htbp]
\centering
\includegraphics[width=0.5\textwidth]{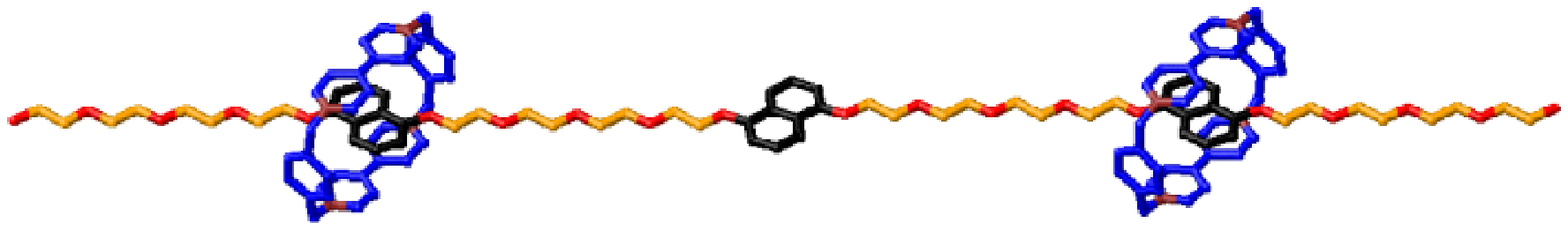} 
\caption{Structure of the [3]rotaxane. Oxygen atoms are depicted in red, the naphthalene units in black, and the polyether carbons in orange.  The \cb rings are depicted in blue with the pyridinium N$^+$ ions in brown. }
     \label{fig:3rotaxane}
\end{figure}

 In this paper we present simulations of  single-molecule pulling for a class of donor-acceptor oligorotaxanes, as well as a detailed interpretation of the observed phenomenology.  The pseudorotaxanes~\cite{stoddartreview95, amabilino} considered consist of a variable number of cyclobis(paraquat-\emph{p}-phenylene) tetracationic cyclophanes (\cb)  threaded onto a linear chain composed of three naphthalene units linked by polyethers, with  polyether caps at each end (see \fig{fig:3rotaxane}). These molecules have been recently  synthesized by Basu \emph{et al.}~\cite{subhadeep}, and are oligomeric analogues of polyrotaxanes previously developed by the Stoddart group~\cite{stoddartpnas}.  The relevant interactions that drive the folding of this class of  molecules have, in the absence of applied stress, been recently characterized~\cite{francoenergetics}.   The interest in this paper is to investigate folding/unfolding pathways for these molecules under stress, and to determine, from a thermodynamic perspective, what drives the folding.  In this sense, the AFM pulling setup is a valuable tool  since  it allows inducing unfolding/refolding events at the single-molecule level while  simultaneously performing thermodynamic measurements.

The AFM pulling of the rotaxanes is simulated using constant temperature MD in a high dielectric medium using  MM3 as the force field~\cite{MM31, MM32, MM33}. We focus on the isometric version of these experiments  in which the distance between the surface and the cantilever is controlled while the force exerted is allowed to vary.  Closely related experimental efforts  are currently  underway in the Stoddart group. To bridge the several orders of magnitude gap between the pulling speeds that are experimentally employed ($1-10^{-3}~\mu$m/s) and those that are computationally accessible, in the simulations the unfolding events are driven slowly enough  that quasistatic behavior is recovered and the results become independent of the pulling speed.

A central quantity of interest in this analysis is the \emph{molecular} potential of mean force (PMF)~\cite{pmf} along the extension coordinate. The PMF is the Helmholtz free energy profile along  a given reaction coordinate.  As such, it succinctly captures the thermodynamic changes experienced by the molecule during the folding.  Here, the PMF is reconstructed  from the equilibrium force measurements  using the weighted histogram analysis method (WHAM)~\cite{wham1, wham2, frenkelandsmit}. Related approaches  based on nonequilibrium force measurements that exploit the Jarzynski~\cite{jarzynski1, jarzynski2, liphardt02} and related equalities  have  been presented previously~\cite{hummer01, hummer05, park04, kiang, imparato08}. Further insight into the thermodynamic driving force behind the folding is obtained by decomposing the PMF  into energetic and entropic contributions.

In addition, an analogy between the phenomenology observed during the pulling of the rotaxanes and that expected for a system undergoing a first-order phase transition is presented. In the process of exploring such an analogy,  several general features of the isometric pulling experiments will be revealed. Specifically, the dependence of the force vs. extension curves on the cantilever spring constant is characterized, and general conditions on the molecule and on the cantilever necessary for the emergence of mechanical instability and blinks  in the force measurements are isolated. In particular, it is shown that the mechanical stability properties observed during the pulling are intimately related to the fluctuations in the force measurements. These results complement previous studies done by Kreuzer \emph{et al.}~\cite{kreuzer2001} in which the soft-spring and stiff-spring limit of the AFM pulling were identified with the Gibbs and Helmholtz ensembles for the isolated molecule,  those of Kirmizialtin \emph{et al.}~\cite{kirmizialtin2005}   in which the origin of the dynamical bistability observed in constant force experiments~\cite{Liphardt01} and its relationship with  the topography of the PMF was clarified, and those of  Friddle \emph{et al.}~\cite{friddle} in which the dependence of the rupture force on the cantilever spring constant was analyzed for a simple model potential meant to represent  bond-breaking (see Ref.~\onlinecite{freund} for an analysis of the bond survival probability). The conditions isolated below  apply to any system subject to equilibrium isometric pulling using AFM tips attached to harmonic cantilevers.

This manuscript is organized as follows.   In \stn{stn:pulling} the protocol employed to simulate the isometric pulling of the rotaxanes is described. Sections \ref{stn:wham}  and \ref{stn:decomposition} summarize, respectively,  the strategy used to reconstruct the PMF from the force measurements and the procedure employed to  decompose it into energetic and entropic  contributions. Our main results are presented in \stn{stn:results}. Specifically,  in \stn{stn:phenomenology} the basic features observed during the mechanical unfolding of the oligorotaxanes are presented  and the resulting  PMF discussed. In \stn{stn:interpretation} the origin of the mechanical instabilities and blinks in the force measurements observed during the extension is clarified, and minimum conditions for the emergence of these effects are presented.  Our main conclusions are summarized in \stn{stn:conclusions}.

\section{Theoretical Methodology}
\label{stn:methodology}

\subsection{Pulling simulations}
\label{stn:pulling}

\begin{figure}[htbp]
\centering
\includegraphics[width=0.3\textwidth]{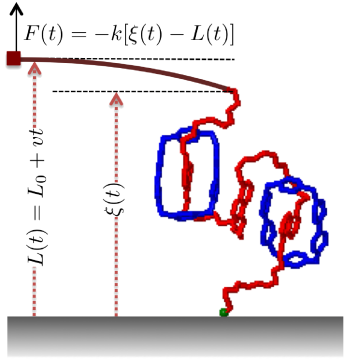} 
\caption{Schematic of an isometric single-molecule force spectroscopy experiment using an atomic force microscope (AFM). In it, one end of the molecular system is attached to a surface and the other end to an AFM tip attached to a cantilever. During the pulling,  the distance between the surface and the cantilever $L(t)=L_0 + vt$ is controlled and  varied at a constant speed $v$.  The deflection of the cantilever from its equilibrium position measures the instantaneous applied force on the molecule by the cantilever  $F(t) = -k [\xi(t) - L(t)]$, where $k$ is the cantilever spring constant and  $\xi(t)$  the fluctuating molecular end-to-end extension.  }
     \label{fig:pullingfig}
\end{figure}

Unfolding of the rotaxanes is computationally studied by means of pulling MD simulations. The simulations are analogous to single-molecule 
force spectroscopy experiments in which a molecule attached to a surface is  
mechanically unfolded by pulling it with an AFM  tip attached to a cantilever. The general setup of this class of experiments is shown in \fig{fig:pullingfig}.
The stretching computation begins by  attaching one end of the molecule to a stiff isotropic harmonic potential that mimics the molecular attachment to the surface. Simultaneously, the opposite molecular  end is connected  to a dummy atom via a virtual harmonic spring.  The position of the dummy atom is the simulation analogue of  the  cantilever position, and is controlled throughout. In turn,   the  varying deflection of the virtual harmonic spring   measures the force exerted during the pulling.  The stretching is caused by moving the dummy atom away from the molecule at a constant speed. The pulling direction is defined by  the  vector  connecting the two terminal atoms  of the complex.   Since cantilever potentials are typically stiff in the direction perpendicular to the pulling, in the simulations the terminal atom that is being pulled is forced  to move along the pulling direction by introducing appropriate additional harmonic restraining potentials.     This type of simulations does not take into account any effects   that may arise due to the interaction between the molecule and the AFM tip or the surface that cannot be accounted for by simple position restraints, nor any variations in the cantilever position  due to thermal fluctuations or solvent-induced viscous drags.

The potential energy function of the molecule plus restraints is of the form
\be
U_L(\vect{r}) = U_0(\vect{r}) + V_L[\xi(\vect{r}), t],
\ee
where  $U_0(\vect{r})$  is the molecular potential energy plus potential restraints not varied during the simulation,  $\vect{r}$  denotes the position of the $N$ atoms in the macromolecule, and  
\be
\label{eq:cantilever}
V_L[\xi(\vect{r}), t] = \frac{k}{2}\left[\xi(\vect{r}) - L(t)\right]^2
\ee
is the potential due to the cantilever. Here $\xi(\vect{r})$ is the molecular end-to-end distance function,  $L= L_0 + v t$  is the distance from the cantilever to the surface at time $t$, $v$ is the pulling speed, and $k$ the cantilever spring constant.  The force exerted by the cantilever on the molecule at time $t$ is given by
\be
\label{eq:force}
F(t) = - \nabla V_L =  \frac{\partial V_L}{\partial L} = - k\left[\xi(\vect{r}) - L(t)\right],
\ee
where the gradient $\nabla$  is with respect to the $[\xi(\vect{r})- L]$ coordinate.
In writing Eqs.~\eqref{eq:cantilever} and~\eqref{eq:force}  the vector nature of  $v(t)$, $L(t)$, $\xi(\vect{r})$  and $F(t)$ has been obviated since  these quantities are collinear in the current setup.  

The simulations are performed using  Tinker 4.2~\cite{tinker} for which a pulling routine was developed. The stretching is performed using  a  soft cantilever  potential  with spring constant $k=0.011$ N/m $= 1.1$ pN/\AA. Since thermal   fluctuations in the force  depend on $k$ as  $\delta F \sim \sqrt{k/\beta}$~\cite{smd1} (where $\beta=1/k_\ti{B} T$ is the inverse temperature), such a soft spring constant provides high resolution  in the force measurements. 
The macromolecules are described using the MM3 force field, which we have found adequately reproduces the complexation energies and the interactions responsible for the folding of the oligorotaxanes~\cite{francoenergetics}. As a simple model for the solvent, we use  a continuum high dielectric medium in which the  overall dielectric constant of the force field is set to that of water at room temperature (78.3). This is to be compared with the MM3 dielectric constant for vacuum of 1.5~\cite{MM34}. Any solvent effects that cannot be described using this continuum description are absent.  
The dynamics is propagated using a  modified Beeman algorithm with a 1 fs integration time step, and the system is coupled to a heat bath at 300 K using  a Nos\'e-Hoover chain as the thermostat.  Initial --minimum energy-- structures  for the pulling simulations were obtained using  simulated annealing,  as described in Ref.~\onlinecite{francoenergetics}. These initial structures were allowed to  equilibrate thermally for  1 ns and subsequently stretched. The simulations do not include  counterions for the tetracationic cyclophanes    since, at 300K and for the  high dielectric medium employed,  their interaction with the main molecular backbone is small and their effect on the folding negligible.

\subsection{Reconstructing the PMF using WHAM}
\label{stn:wham}

Under reversible conditions, knowledge of the force exerted during the  pulling immediately yields the associated change  in the Helmholtz free energy $A$ for the molecule plus cantilever. This is because  the change in $A$ is determined by the reversible work exerted during the pulling
\be
\label{eq:helmholtz}
\Delta A   = \int_{L_0}^{L}  \langle F \rangle_{L'}  \, \ud L' = -\frac{1}{\beta}\ln\frac{Z(L)}{Z(L_0)},
\ee
where  $\langle F \rangle_L$ is the average force at extension $L$, and 
\be
\label{eq:partfuncL}
Z(L) = \int \ud\vect{r} \exp[-\beta U_L(\vect{r})]
\ee
is  the configurational partition function of the molecule plus cantilever. Equation~\eqref{eq:helmholtz} follows from standard thermodynamic integration considerations~\cite{frenkelandsmit}, and  assumes that at each point during the extension the state of the molecule plus cantilever potential  is well described by a canonical ensemble. This property is expected for a system in weak contact with a thermal bath and  is  enforced in the simulations by the nature of the thermostat employed.

 From the force versus extension measurements  it is also  possible to extract the molecular potential of mean force (PMF)   $\phi(\xi)$  as a function of the  end-to-end distance  $\xi$ by properly removing the bias due to the cantilever potential. The quantity $\phi(\xi)$ is the Helmholtz free energy profile along the  coordinate $\xi$ for the isolated (cantilever-free) molecule, and  is of central  interest since it succinctly characterizes the thermodynamics of the unfolding.  
Below we  summarize  how to estimate this quantity from the force measurements employing the weighted histogram analysis method (WHAM). 
  
 The PMF is defined by~\cite{whamroux}
\be
\label{eq:PMF}
\phi(\xi)  = \phi(\xi^\star) -\frac{1}{\beta}\ln \left[ \frac{p_0(\xi)}{p_0(\xi^\star)} \right],
\ee
 where $\xi^\star$ and  $\phi(\xi^\star)$ are arbitrary constants and 
\be
\label{eq:unbiased}
 p_0(\xi) = \frac{\int \ud \vect{r}\,  \delta[\xi-\xi(\vect{r})] \exp[-\beta U_0(\vect{r})]}{ \int \ud \vect{r} \,\exp[-\beta  U_0(\vect{r}) ] }  = \frac{Z_0(\xi)}{Z_0} = \langle\delta[\xi- \xi(\vect{r})]\rangle.
\ee
Here, 
\be
Z_0 = \int \ud \vect{r} \,\exp[-\beta  U_0(\vect{r}) ]
\ee
 is the  configurational partition function of the molecule, and
\be
\label{eq:partdens}
Z_0(\xi) = \int \ud \vect{r}\,  \delta[\xi-\xi(\vect{r})] \exp[-\beta U_0(\vect{r})].
\ee
In the context of the pulling experiments  $p_0(\xi)$ is the probability density that the molecular end-to-end distance function  $\xi(\vect{r})$  adopts the value $\xi$ in the unbiased (cantilever-free) ensemble.  The quantity   $p_0(\xi)$ determines the PMF  up to a  constant and can be estimated from the force measurements, as we now describe.

At this point it is  convenient to discretize the time variable during the pulling experiment  into $M$ steps:  $ t_1, \ldots, t_i, \ldots,  t_M$. The potential  of the system plus cantilever at time $t_i$ is   $U_i = U_0 + V_i$, where $V_i = \frac{k}{2}[\xi(\vect{r}, t_i) - L(t_i)]^2$ is the bias due to the cantilever potential. In the $i$-th biased measurement knowledge of the force $F_i$ and  $L(t_i)$ gives the molecular end-to-end distance.  The probability density of observing the value  $\xi$  at this time  is given by
\be
\label{eq:biased}
p_i(\xi) = \frac{\int \ud \vect{r}\, \delta[\xi-\xi(\vect{r})] \exp[-\beta U_i(\vect{r})]    }{\int \ud \vect{r}  \, \exp[-\beta U_i(\vect{r})]  } =  \frac{Z_i(\xi)}{Z_i} = \langle \delta[ \xi - \xi(\vect{r})] \rangle_i,
\ee
where  $Z_i= \int \ud \vect{r}\,  \exp[-\beta U_i(\vect{r})]$ is the configurational partition function for the system plus cantilever at the $i$-th extension, and
\be
Z_i(\xi) = \int \ud \vect{r} \delta[\xi-\xi(\vect{r})] \exp[-\beta U_i(\vect{r})].
\ee
Knowledge of $p_i(\xi)$  allows  the unbiased probability density $p_0(\xi)$ to be reconstructed since, by virtue of Eqs.~\eqref{eq:unbiased} and~\eqref{eq:biased}, these two quantities are  related by 
\be
p_0(\xi)  =  \exp[ + \beta V_i(\xi)] \frac{Z_i}{Z_0} p_i(\xi),
\ee
where we have exploited the fact that  $Z_0(\xi)= \exp[+\beta V_i(\xi)]Z_i(\xi)$.
In the pulling experiments, for each $L(t_i)$ only a few force measurements are typically performed  and the probability density  $p_i(\xi)$ is not  properly sampled. Nevertheless, the experiments  do provide  measurements at a wealth of  values of $L(t_i)$ that can be combined  to estimate  $p_0(\xi)$:
\be 
p_0(\xi) = \sum_{i=1}^{M} w_i \exp[+\beta V_i(\xi)]\frac{Z_i}{Z_0} p_i(\xi),
\ee
where the $w_i$ are some normalized ($\sum_i w_i = 1$) set of weights. In the limit of perfect sampling any set of weights should yield  the same $p_0(\xi)$. In practice, for finite sampling it is convenient to employ a set that minimizes the  variance in the  $p_0(\xi)$ estimate  from the series of independent estimates of biased distributions. Such a set of weights  is precisely provided by the WHAM prescription~\cite{wham1, wham2, frenkelandsmit,wham3}
\be
\label{eq:weights}
w_i  = \frac{\exp[-\beta V_i(\xi)] \dfrac{Z_0}{Z_i g_i}}{\sum_{i=1}^{M}  \exp[-\beta V_i(\xi)] \dfrac{Z_0}{Z_i g_i}},
\ee
where $g_i = 1+2\tau_i$ is the statistical inefficiency~\cite{wham3}  and  $\tau_i$   the integrated autocorrelation time~\cite{wham1, wham3}. The difficulty in estimating the $\tau_i$ for each measurement generally leads to further supposing that the $g_i$'s are approximately constant  and  factor out of \eq{eq:weights}, so that
\be
\label{eq:wham}
p_0(\xi)  = 
\frac{\sum_{i=1}^{M} p_i(\xi)}{\sum_{i=1}^{M}  \exp[-\beta V_i(\xi)] Z_0/Z_i}.
\ee
Neglecting the $g_i$'s from \eq{eq:wham} does  not imply that the resulting estimate of  $p_0(\xi)$ is incorrect; but simply that  the weights selected do not precisely minimize the variance in the estimate. Experience with this method indicates that  if the $g_i$'s do not differ by more than an order of magnitude their effect on $\phi(\xi)$ is small~\cite{wham2}.

The computation of the PMF then proceeds as follows. Suppose that $N_i$ force measurements are done for each extension  $i$.   Knowledge of the force $F_i^j$ ($j=1, \ldots, N_i$) and of $L(t_i)$ gives the molecular end-to-end distance $\xi_i^j$ in each of these measurements.  The numerator in \eq{eq:wham} is then estimated by  constructing a histogram with all the available data,
\be
\sum_{i=1}^{M} p_i(\xi) \approx \sum_{i=1}^{M}\sum_{j=1}^{N_i} \frac{C_i^{j}(\xi)}{N_i \Delta\xi},
\ee
where $\Delta\xi$  is the bin size,  and $C_i^j(\xi)= 1$ if $\xi_i^j\in [\xi - \Delta\xi/2, \xi + \Delta\xi/2)$ and zero otherwise.  Estimating the  denominator in \eq{eq:wham} requires knowledge of the $Z_i$'s, the configurational partition functions of the system plus cantilever at all extensions $\{L(t_i)\}$ considered. There are two ways to obtain these quantities. The most direct one is to employ the Helmholtz free energies for the system plus cantilever obtained through the thermodynamic integration in \eq{eq:helmholtz}, as they determine  the $Z_i$'s  up to a constant multiplicative factor. Alternatively,  it is also possible to determine the ratio of the configurational partition functions between  the $i$-th biased system and its  unbiased counterpart using $p_0(\xi)$:
\be
\label{eq:partitionf}
\frac{Z_i}{Z_0}  = \int \ud\xi\, \exp[-\beta V_i(\xi)] p_0(\xi).
\ee
Equations~\eqref{eq:wham} and~\eqref{eq:partitionf}  can be solved iteratively. Starting from a guess for the $Z_i/Z_0$, $p_0(\xi)$ is estimated using \eq{eq:wham} and normalized. The resulting $p_0(\xi)$ is then used to obtain a new set of $Z_i/Z_0$ through \eq{eq:partitionf}, and the process is repeated until self-consistency. This latter approach is the usual procedure to solve the WHAM equations. Note, however, that this procedure is not required  if reversible force vs. extension data is available.

The reconstruction of the PMF from the force vs. extension measurements using WHAM is simple to implement. Further, the procedure is  versatile in the sense that it makes no assumption about the stiffness of the cantilever potential and  permits the combination of data from several pulling runs. In addition,  as we describe below, it can also be employed to decompose the PMF into entropic and energetic contributions from data readily available from an MD run.

\subsection{Energy-entropy decomposition of the PMF}
\label{stn:decomposition}

Insight into the thermodynamic driving force responsible for the folding of oligorotaxanes can be obtained by  decomposing  the changes in the PMF  into entropic $S(\xi)$  and potential energy $U_0(\xi)$ contributions,
\be
\label{eq:decomposition}
\phi(\xi) -\phi(\xi^*)  =   \left[U_0(\xi) -U_0(\xi^\star)\right] - T\left[S(\xi) -S(\xi^*) \right].
\ee
Here, 
\be
 U_0(\xi)  = \frac{1}{Z_0(\xi)} \int \ud\vect{r}\, \delta[\xi-\xi(\vect{r})]U_0(\vect{r}) \exp[-\beta U_0(\vect{r})]
\ee
 is the average potential energy when the end-to-end distance adopts the value $\xi$, and 
\be
TS(\xi) = -\frac{1}{\beta \Omega_0} \int \ud\vect{r}\, \delta[\xi-\xi(\vect{r})] \rho \ln \rho
\ee
is the corresponding configurational entropy, where  $\rho  = \frac{\Omega_0}{Z_0(\xi)} \exp[-\beta U_0(\vect{r})]$ and $\Omega_0$ is a constant irrelevant factor with the same dimensions as $Z_0(\xi)$.
Equation~\eqref{eq:decomposition} follows  from the above definitions. 

The energy-entropy decomposition of the PMF  can be obtained by estimating   $ U_0(\xi)$ directly and employing the previously reconstructed $\phi(\xi)$ to obtain $TS(\xi)$ using \eq{eq:decomposition}. Here, $ U_0(\xi)$ is estimated  by  combining all measurements  performed during the pulling simulations using WHAM.  The measurable quantity  in the simulation in the presence of the bias due to the cantilever  is:
\be
\langle U_0(\vect{r}) \delta[\xi-\xi(\vect{r})]   \rangle_i =
\frac{\int \ud\vect{r} \,\delta[\xi-\xi(\vect{r})]U_0(\vect{r}) \exp[-\beta U_i(\vect{r})]}{Z_i},
\ee 
the  average molecular potential energy when the end-to-end distance adopts the  value $\xi$ and the cantilever extension is $L(t_i)$.  This average in the biased ensemble is related to  the unbiased average $U_0(\xi)$  by:
\be
 U_0(\xi)  = \exp[+\beta V_i(\xi)]\frac{Z_i}{Z_0(\xi)} \langle U_0(\vect{r}) \delta[\xi-\xi(\vect{r})]   \rangle_i.
\ee
As in the estimation of $p_0(\xi)$, all data collected during  pulling can be combined with appropriate weights $w_i$ to estimate $U_0(\xi)$, 
\be
 U_0(\xi)  = \sum_{i=1}^{M} w_i \exp[+\beta V_i(\xi)]\frac{Z_i}{Z_0(\xi)} \langle U_0(\vect{r}) \delta[\xi-\xi(\vect{r})]   \rangle_i.
\ee
If one adopts the weights $w_i$ that define the  WHAM prescription [\eq{eq:weights}], then 
\be
U_0(\xi)  =  \frac{\sum_i \langle U_0(\vect{r}) \delta[\xi-\xi(\vect{r})]   \rangle_i}{\sum_i  \exp[-\beta V_i(\xi)] Z_0(\xi)/Z_i },
\ee
where we have assumed identical  statistical inefficiencies for all $i$.   Last, recalling  the definition of $p_0(\xi)= Z_0(\xi)/Z_0$ [\eq{eq:unbiased}] and its WHAM prescription [\eq{eq:wham}],  one arrives at a useful expression for $U_0(\xi)$, 
\be
\label{eq:potwham}
 U_0(\xi)   =  \frac{\sum_i \langle U_0(\vect{r}) \delta[\xi-\xi(\vect{r})]   \rangle_i}{\sum_i  \langle \delta [\xi-\xi(\vect{r})] \rangle_i}.
\ee
Equation~\eqref{eq:potwham}  provides  a mean to calculate the potential energy as a function of the end-to-end distance  by combining all the data obtained during the pulling.  The remarkable simplicity of \eq{eq:potwham} is noteworthy:  in order to obtain $U_0(\xi)$ one just needs to   generate histograms for the the values of $\xi$ observed using all data collected during the pulling and then, within each bin, perform a simple average of the internal energy of the configurations that fall into it.  It does not require knowledge of the properties of the bias potential. 

This method of performing an energy-entropy decomposition of the PMF requires knowledge of  $U_0(\xi)$ for the configurations observed during pulling. This quantity, although readily available in an MD run, is not experimentally accessible.  As a result, in experiments  an energy-entropy decomposition of the PMF requires estimating the changes in entropy by performing the pulling at  varying temperatures.

\section{Results and discussion}
\label{stn:results}

\subsection{Phenomenology of the pulling experiments}
\label{stn:phenomenology}

\subsubsection{The approach to equilibrium}
\label{stn:toeql}

One of the challenges in simulating AFM pulling experiments  using MD is to bridge the  large  disparity between the pulling speeds $v$ that are  employed  experimentally and those that can be accessed computationally.   While typical experiments often use   $v \sim 1-10^{-3}$ $\mu$m/s, current computational capabilities require pulling speeds that  are several orders  of magnitude faster.  Here, this difficulty is circumvented by striving for  pulling speeds that are slow enough that   reversible behavior is recovered. Under such conditions   the results become independent of $v$, and the simulations comparable to experimental findings.

\begin{figure}[htbp]
\centering
\includegraphics[width=1.0\textwidth]{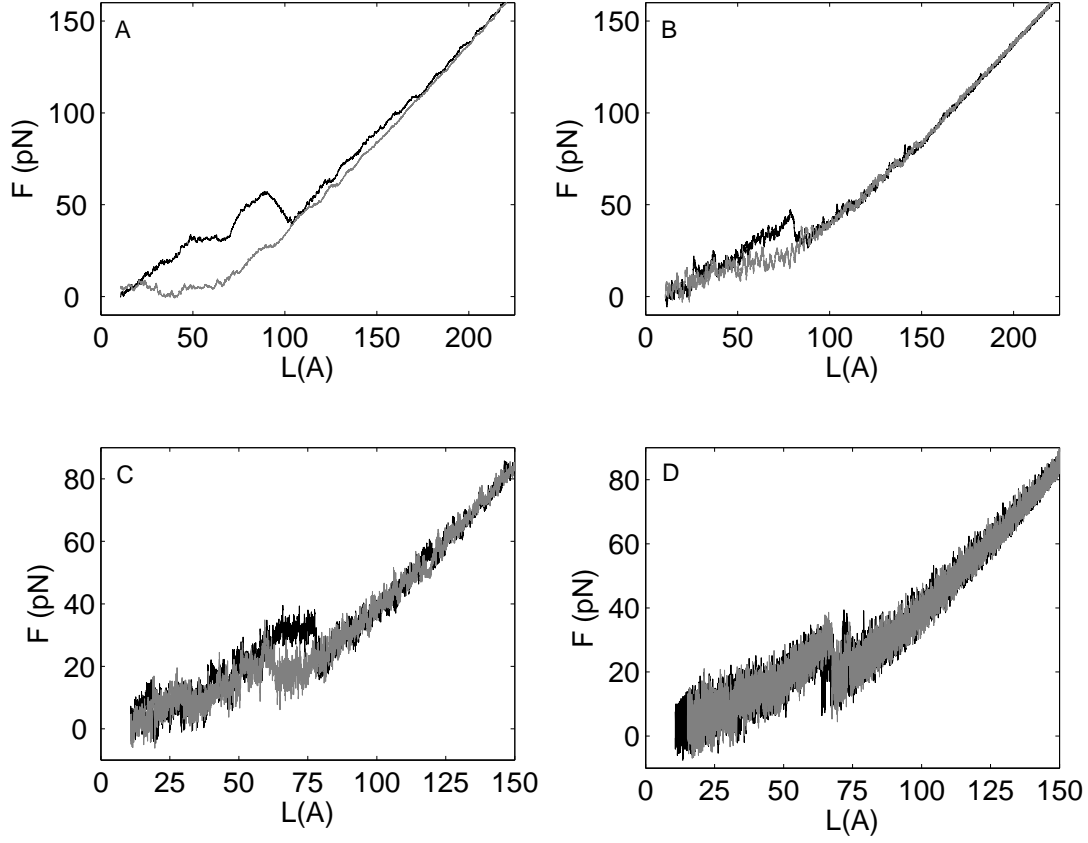} 
\caption{The approach to equilibrium in the pulling simulations. The figure shows force  vs. extension  profiles  for the [3]rotaxane  immersed in a high dielectric medium at 300K  for different pulling speeds $v$ of (A) $10^0$, (B) $10^{-1}$, (C)  $10^{-2}$  and (D) $10^{-3}$ \AA/ps. The  harmonic cantilever employed has a soft spring constant of  $k=0.011$ N/m, $L$ is the distance between the surface and the cantilever and $F$ the instantaneous  applied force.   The black  lines correspond to the extension process; the gray ones to the contraction.  }
     \label{fig:3DNP2ring.approachtoeql}
\end{figure}

Consider the pulling of the [3]rotaxane shown in \fig{fig:3rotaxane} (the simulation details are specified in \stn{stn:pulling}).  Figure~\ref{fig:3DNP2ring.approachtoeql} shows the effect of decreasing the pulling speed on the force vs. extension  characteristics. In the simulations, the system is first extended  (black lines) to a given $L$ and then contracted (gray lines).    For pulling speeds (and equal retracting speeds) between $1-0.01$ \AA/ps (panels A to C) hysteresis and other nonequilibrium effects during the pulling are unavoidable. However, for $v=0.001$ \AA/ps  $=10^5$ $\mu$m/s  the system behaves reversibly, and the  $F$-$L$ curves obtained during extension and contraction essentially coincide. Note the  second law of thermodynamics at play in the simulations: the work required to stretch the molecule --the area under the curve-- decreases with the pulling speed until reversible behavior is attained. We have observed that   pulling speeds of $\sim10^{-3}$ \AA/ps are generally sufficient to recover reversible  behavior  for the unfolding in a high dielectric medium of the [n]rotaxanes considered. For simulations in vacuum slower pulling speeds are required.

It is worth noting  that this  seemingly simple phenomenological behavior  is not recovered when a Berendsen thermostat is employed instead of a Nos\'e-Hoover chain. In fact, we have observed that the Berendsen thermostat leads to a spurious violation of the second law  during  the pulling, presumably because the ensemble generated by it does not  satisfy the equipartition theorem~\cite{flyingicecube}.

\subsubsection{Reversible unfolding, mechanical instability and blinks in the force measurements}
\label{stn:mainfeatures}

\begin{figure}[htbp]
\centering
\includegraphics[width=1.0\textwidth]{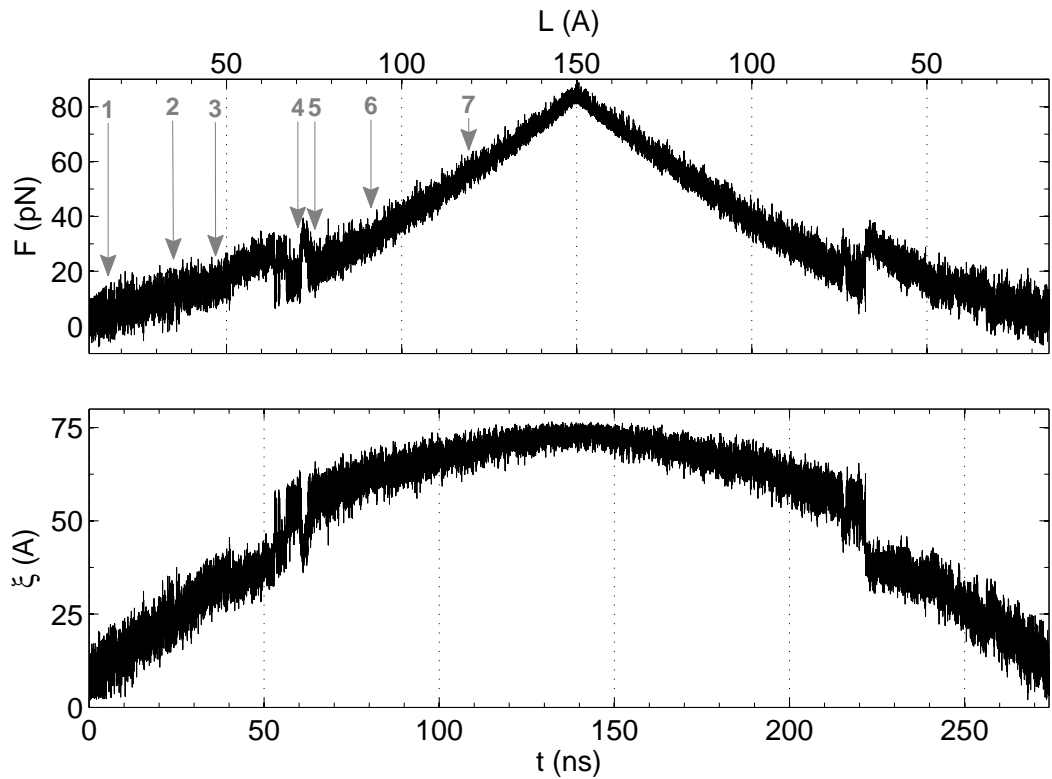} 
\caption{Time dependence of the force $F$ and the  molecular end-to-end distance $\xi$  during the pulling of the [3]rotaxane under equilibrium conditions ($v= 0.001$ \AA/ps, $k=0.011$ N/m). Typical structures (labels 1-7) observed during the extension are shown in \fig{fig:3DNP2ringsnapshots}. }
     \label{fig:3DNP2ring.equilibrium}
\end{figure}

\begin{figure}[htbp]
\centering
\includegraphics[width=0.6\textwidth]{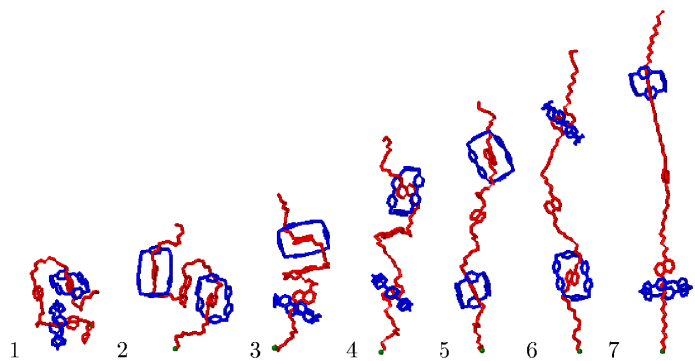}
\caption{Snapshots of the [3]rotaxane during its extension.  The numerical labels shown here are employed in Figs.~\ref{fig:3DNP2ring.equilibrium},\ref{fig:3DNP2ring.freeenergysc} and \ref{fig:3DNP2ring.molecularphi}. }
     \label{fig:3DNP2ringsnapshots}
\end{figure}

Figure~\ref{fig:3DNP2ring.equilibrium} details the force exerted (upper panel) and the molecular end-to-end distance (lower panel)  during the reversible  pulling and  contraction of the [3]rotaxane.  Snapshots of  typical structures encountered during the pulling are included in \fig{fig:3DNP2ringsnapshots}. As the system is stretched the oligorotaxane undergoes a conformational transition from a folded globular state to an extended coil. In the process, the force initially  increases approximately linearly with $L$,  then  drops, and subsequently increases again.  The drop in the force is due to the unfolding of the rotaxane during the stretching.

Figure~\ref{fig:3DNP2ring.freeenergysc} shows the change in the Helmholtz free energy $A$ for the molecule plus cantilever obtained from integrating directly the $F$-$L$ curve in \fig{fig:3DNP2ring.equilibrium}.  The net change in free energy for the complete thermodynamic cycle is zero, as  expected for a quasistatic process.  The free energy profile for the system plus cantilever has three distinct regions, labeled I-III in the figure.  Regions I  and III correspond to the folded and extended state, respectively, while region II is where the folding/unfolding event occurs. From a thermodynamics perspective~\cite{callen}, regions I and III are mechanically stable phases  since $A$ is a convex function of the extension $L$, i.e. $\left(\frac{\partial^2 A}{\partial L^2}\right)_T = \left( \frac{\partial \langle F \rangle_L}{\partial L}\right)_T   \ge 0$.   By contrast, region  II, where the unfolding occurs, is mechanically unstable since $A$ is a concave function of $L$ and hence $\left(\frac{\partial^2 A}{\partial L^2}\right)_T = \left( \frac{\partial \langle F \rangle_L}{\partial L}\right)_T  < 0$.

The dynamical behavior of the system in the unstable region is illustrated in \fig{fig:3DNP2ring.unstable}, which shows the evolution of the radius of gyration $R_g$ and  the molecular end-to-end distance when $L=70.0$~\AA.    In this region,  the  [3]rotaxane undergoes transitions between  a  folded globular state, partially folded structures, and  an extended coil.  The right panels in  \fig{fig:3DNP2ring.unstable} show the  probability density of the distribution of  $R_g^2$ and  $\xi$ values obtained from a 20 ns trajectory. The system exhibits a clear dynamical bistability along the $\xi$ coordinate, and at  least a   tristability  along $R_g^2$. Figure~\ref{fig:3DNP2ring.unstable.structures}  shows some representative structures encountered during this dynamics. The bistability along the end-to-end distance  leads to a blinking in the force measurements from a high force to a low force regime during the pulling (cf. the unstable region in \fig{fig:3DNP2ring.equilibrium}).   The blinking in the force measurements in the isometric experiments is analogous to the blinking in the molecular extension observed during the constant force reversible pulling of single RNA molecules~\cite{Liphardt01}. Note that for the cantilever stiffness employed this multistable behavior in the radius of gyration, or the end-to-end distance, is  not observed when $L$ is set to be in the stable regions (I or III) of the free energy profile.   
  The origin of the bistability along the end-to-end distance, the multistability in the radius of gyration,   and its relation to the thermodynamic instability in the pulling is discussed in \stn{stn:interpretation}.

\subsubsection{The PMF and the thermodynamic driving force in the folding}
\label{stn:drivingforce}

 Figure~\ref{fig:3DNP2ring.molecularphi} shows the potential of mean force along the end-to-end distance $\xi$ for the [3]rotaxane reconstructed from the force measurements as described in \stn{stn:wham}, as well as its decomposition into entropic and potential energy contributions.   The thermodynamic native state of the oligorotaxane, i.e. the minimum in $\phi$, corresponds to a  globular folded structure with $\xi_0=7$~\AA. The structure of the molecular free energy profile as a function of $\xi$ is similar to that observed for the molecule plus cantilever as a function of $L$. Namely, it consists of a  folded and an unfolded stable branch where $\phi(\xi)$ is a convex function,  and a concave region around $\xi= 44-52$ \AA~ where the conformational transition occurs. As discussed in \stn{stn:interpretation}, this characteristic structure of the PMF is responsible for many of the interesting features observed during pulling.  Figure~\ref{fig:unstablenocantilever} shows the probability distribution for  $R_g^2$ at values of $\xi$ in the three main regions of the PMF,  each collected during  a 20 ns trajectory. In the stable regions ($\xi=7.0$ and 64.9 \AA~), the evolution of the radius of gyration reveals only one stable structural state. By contrast, in the unstable region  ($\xi=47$ \AA) the distribution of $R_g^2$ values  indicates that there is a inherent bistability in the molecular potential along the $R_g^2$ coordinate.  Such molecular bistability gives rise to  the region of concavity in the PMF.

What drives the folding of the oligorotaxanes from a thermodynamic perspective? The energy-entropy decomposition of the PMF indicates that the folding  of the rotaxane is energetically favored but entropically penalized, with the energetic contributions overcoming the entropy penalty and effectively driving the self-assembly.   Further, while the potential energy and the entropy of the molecule show considerable changes during the unfolding, these two effects largely cancel one another, leading to only modest changes in the  Helmholtz free energy.  Note that while  $\phi(\xi)$ is a monotonic function of $\xi$ for $\xi> \xi_0$, the  energy and entropy contributions show a clear nonmonotonic dependence on $\xi$. For instance,  there is a reduction in the system's entropy for $\xi$ both just before the conformational transition and for large extensions due to a reduction in the available degenerate conformational  space. 

It is important to stress that the PMF and its decomposition into energy and entropy contributions correspond to a molecule in which the terminal atoms are constrained  to move along the pulling coordinate. As a consequence, any additional  contributions that may arise by relaxing these constraints are simply not manifest in this picture of the folding.

\subsubsection{Dependence of the PMF on the number of threaded rings}
\label{stn:norings}

\begin{figure}[htbp]
\centering
\includegraphics[width=0.6\textwidth]{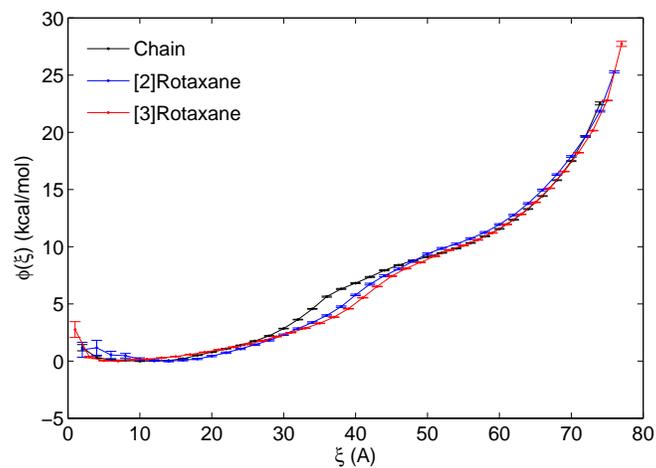}
\caption{PMF as a function of the molecular extension $\xi$ extracted from force measurements using WHAM for rotaxanes with different numbers of threaded rings.  The error bars correspond to twice the standard deviation obtained from a bootstrapping analysis.   }. 
     \label{fig:noofrings}
\end{figure}

Figure~\ref{fig:noofrings} compares the PMF  extracted from force measurements for the [3]rotaxane, and the associated   [2]rotaxane and bare molecular thread.  In the [2]rotaxane system the \cb ring encircles the central naphthalene unit of the underlying chain.  The PMF in all cases is qualitatively similar: it consists of two stable convex regions,  and a unstable concave  region where the unfolding occurs. Further, for all species the folding along $\xi$ is energetically driven and entropically penalized (not shown).  Note that, in the high dielectric medium employed, the number of threaded rings does not have a dramatic effect on the net change in the free energy undergone during folding nor, as expected, on the elastic properties of the extended state. It does, however, have an appreciable effect on the stability and elasticity of the folded conformation. Specifically, as the number of rings is increased the range of $\xi$ values for which the folded conformation is stable increases. That is, a larger $L$ is required during the pulling to unfold the rotaxanes with respect to the bare chain.

\subsection{Interpretation of the pulling experiments}
\label{stn:interpretation}

\subsubsection{Analogy with first-order phase transitions}
\label{stn:phasetransition}

\begin{figure}[htbp]
\centering
\includegraphics[width=0.7\textwidth]{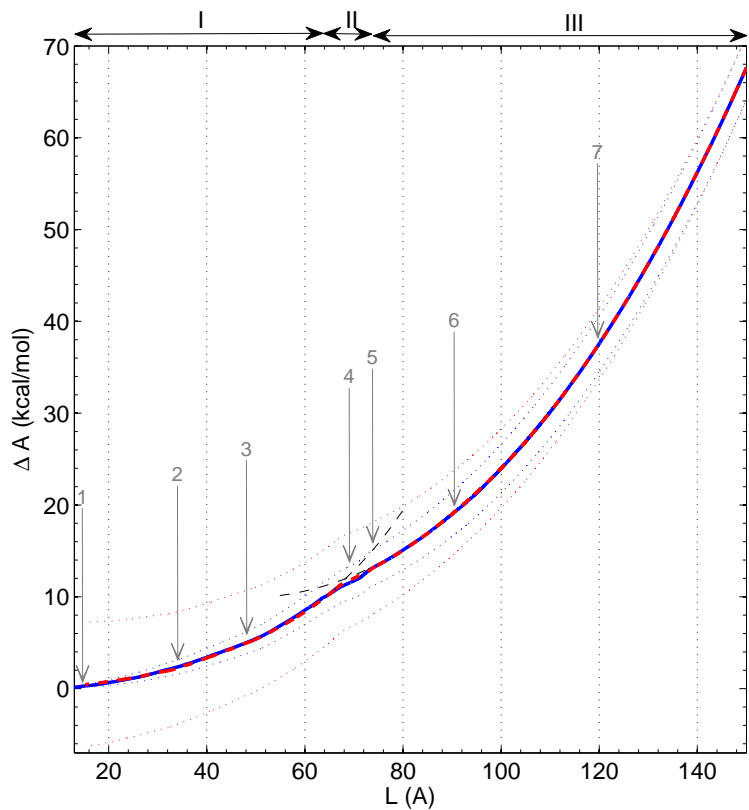} 
\caption{Changes in the Helmholtz free energy $A$ of the [3]rotaxane plus cantilever obtained from the data shown in  \fig{fig:3DNP2ring.equilibrium}. The blue line corresponds to the pulling and the red one to the contraction.    The dotted lines provide an estimate of errors in the thermodynamic integration due to force fluctuations at each pulling step. The difference in the degree of convexity of regions I and III  is evidenced by extrapolating the data in each region outside of its domain through  fitting to  quartic polynomials (black dashed lines).  The labels correspond to the structures shown  in \fig{fig:3DNP2ringsnapshots}.  }
     \label{fig:3DNP2ring.freeenergysc}
\end{figure}

\begin{figure}[htbp]
\centering
\includegraphics[width=0.8\textwidth]{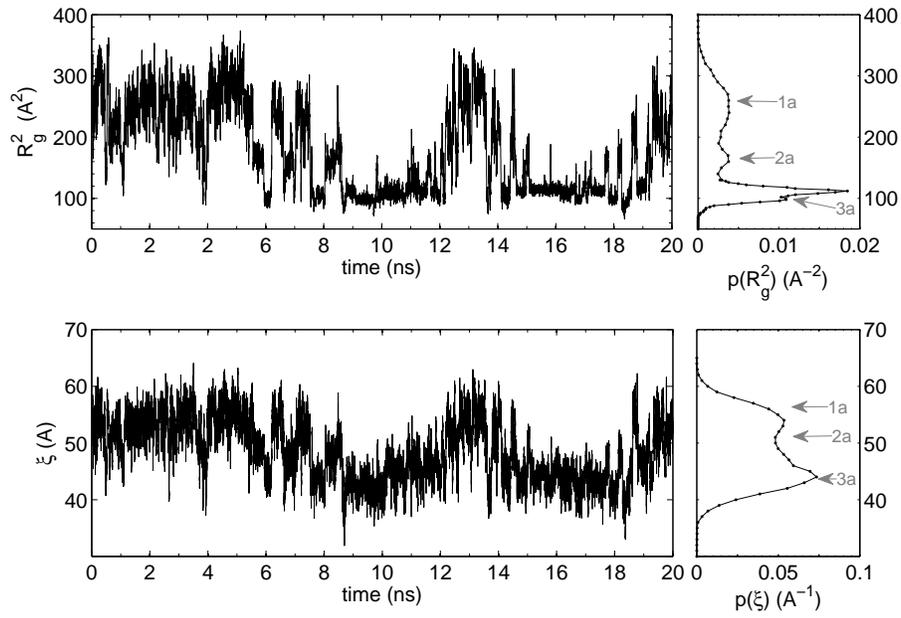} 
\caption{Time dependence and probability density distribution  of the radius of gyration ($R_g^2$) and the end-to-end molecular extension $\xi$ when the [3]rotaxane plus cantilever is constrained to reside in the unstable region (II)  of \fig{fig:3DNP2ring.freeenergysc}, with $L=70.0$ \AA. Typical structures encountered in this regime (labels 1a-3a) are  shown in \fig{fig:3DNP2ring.unstable.structures}. }
     \label{fig:3DNP2ring.unstable}
\end{figure}

\begin{figure}[htbp]
\centering
\includegraphics[width=0.3\textwidth]{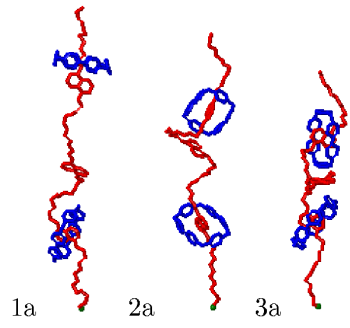}
\caption{Structures observed in the unstable region of the pulling simulations.  The numerical labels are employed in \fig{fig:3DNP2ring.unstable}.  }
     \label{fig:3DNP2ring.unstable.structures}
\end{figure}

At first glance, the phenomenological behavior observed by the molecule plus cantilever during the extension   is analogous to that expected for a system undergoing a first-order phase transition~\cite{callen}.  Namely, during the extension the combined system goes from one stable thermodynamic phase to another (Regions I and III in \fig{fig:3DNP2ring.freeenergysc}) by varying an externally controllable parameter, in this case  $L$. In the transition region the Helmholtz free energy of the combined system  is concave and, hence, fails to satisfy the thermodynamic stability criteria (Region II in \fig{fig:3DNP2ring.freeenergysc}). As a consequence, the observed  $F$-$L$ isotherm (\fig{fig:3DNP2ring.equilibrium})  exhibits a region of mechanical instability where  $\left(\frac{\partial \langle F \rangle_L}{\partial L} \right)_T < 0$.  These observations  largely parallel the behavior of unstable $P$-$V$ isotherms for a van der Waals fluid.

In addition, the phenomenology  seems to indicate that close to the transition region the Helmholtz potential for the molecule plus cantilever is bistable along the end-to-end distance, and has the characteristic form exhibited during a first-order phase transition schematically shown  in \fig{fig:phasetransition}. For  $L\in$  Region I  the stable global  minimum corresponds to the molecule in a folded state, and as $L$ is increased the equilibrium state shifts from one local minimum to the other. For $L\in$ Region II the two minima become approximately equal and a transition from the folded state to the unfolded state occurs. Beyond this extension the unfolded  molecular phase becomes the global potential minimum and the extended phase becomes absolutely stable.  This interpretation is consistent with the observed behavior in which the system blinks between a folded and an extended state  around the transition region, recall \fig{fig:3DNP2ring.unstable}. 

Note, however, that this analogy can never be completely faithful since the modeled  single-molecule  pulling experiment does not deal  with a macroscopic system. This contrasts with the elastic behavior observed for polymer chains~\cite{halperin91, walker07}, and leads to  salient differences between the two processes. First,  while in a true first-order phase transition the physical system exhibits coexistence between the stable phases, in this single-molecule version no coexistence is possible. At the transition point, and for the value of $k$ employed, the system instead exhibits frequent jumps between the stable phases (cf. \fig{fig:3DNP2ring.unstable}). Coexistence is replaced by ergodicity in this single-molecule manifestation of bistability.   Second,  the variable that is varied during the extension $L$  is not a true thermodynamic variable since it is neither extensive nor intensive. Third, the system considered has no clear thermodynamic limit in which a true discontinuity of the derivatives  of the partition function can develop and, further,  the system  is far from the thermodynamic limit since the fluctuations observed in the $F$-$L$ curves are comparable to the average value, leading to a nonequivalence between statistical ensembles~\cite{hill, rubi}.  Additionally,  the mechanical properties  measured during the pulling are for the molecule plus cantilever, and hence the extension behavior is expected to depend on the nature of the cantilever. 

Nevertheless, the similarities between the two processes are still intriguing and, in view of these observations, it is natural to ask: (i) what is the origin of the mechanical instability during the pulling?; (ii) how does the dynamical bistability along the end-to-end distance arise?;  (iii)  how does the nature of  the cantilever affect the observed phenomenological behavior; and (iv) which features of the observed behavior  arise due to the molecule, and which  due to the cantilever?  These questions are addressed in the following sections.

 The key quantity in the remainder of this analysis is the PMF along the end-to-end distance  [\eq{eq:PMF}]. Its utility relies on the fact that the  configurational partition function of the molecule plus cantilever at extension $L$ [\eq{eq:partfuncL}]  can be expressed as
\be
\label{eq:partfuncL2}
Z(L) = \int\ud {\xi}\, \exp\{-\beta[\phi(\xi) + V_L(\xi)] \}, 
\ee
where we have neglected the  constant and $L$-independent multiplicative factor that arises in the transformation since it is irrelevant for the present purposes. The above relation implies that the extension process can be viewed as thermal motion along  a one-dimensional effective potential determined by the PMF $\phi(\xi)$ and the bias due to the cantilever $V_L(\xi)$:
\be
\label{eq:effectivepotential}
U_L(\xi) = \phi(\xi) + V_L(\xi).
\ee
Further, by means of the PMF  it is  possible to estimate the force vs. extension characteristics for the composite system for  any value of the force constant $k$. This is because  the average force exerted on the system at extension $L$ can be expressed as:
\be
\label{eq:averageforce}
\langle F \rangle_L = - \frac{1}{\beta} \frac{\partial  \ln[Z(L)/Z_0]}{\partial L} 
  = \frac{\int \ud\xi \, \frac{\partial V_L(\xi)}{\partial L}\exp\{ -\beta [\phi(\xi) + V_L(\xi)]\}   }
{ \int \ud\xi\, \exp \{ -\beta [\phi(\xi) + V_L(\xi)]\}   },
\ee
 where, for convenience, we have  introduced $Z_0$ in the logarithm just to make the argument  dimensionless. Since $\phi(\xi)$   is  a property of the isolated molecule it is  independent of $k$. Hence,   \eq{eq:averageforce} can be employed to estimate the $F$-$L$ isotherms for arbitrary $k$.

\subsubsection{Dependence of the $F$-$L$ isotherms on the cantilever spring constant}
\label{stn:effectofk}

\begin{figure}[htbp]
\centering
\includegraphics[width=1.0\textwidth]{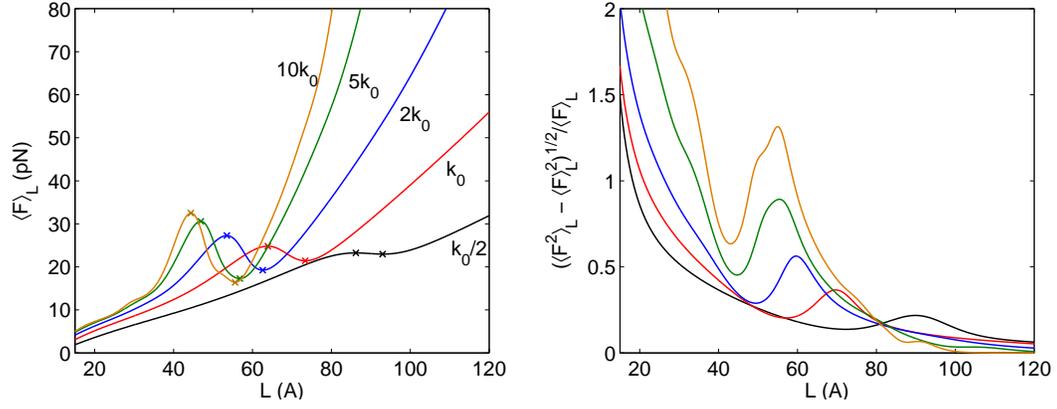}
\caption{Left panel: dependence of the $F$-$L$ isotherms  on the cantilever spring constant $k$ during the extension of the [3]rotaxane. Right panel: ratio between the thermal fluctuations and the average  in the force measurements. The cantilever spring constants employed are all expressed in terms of the cantilever spring constant used in the pulling simulations  $k_0=1.1$ pN/\AA.  } 
     \label{fig:FvsLdifferentk}
\end{figure}

Figure~\ref{fig:FvsLdifferentk} shows representative $F$-$L$ curves  for the [3]rotaxane estimated using \eq{eq:averageforce} for different cantilever spring constants, as well as the  ratio between the thermal  fluctuations and the average in the force measurements. Note that in all cases the system is far from the thermodynamic limit since  the force fluctuations are comparable in magnitude to the average.  Further, the fluctuations increase with the spring constant  and, for example, in the case of $k=5k_0$ (where $k_0=1.1$ pN/\AA) the region of instability  in the $F$-$L$ curve would be   mostly masked by the fluctuations.

\begin{figure}[htbp]
\centering
\includegraphics[width=0.9\textwidth]{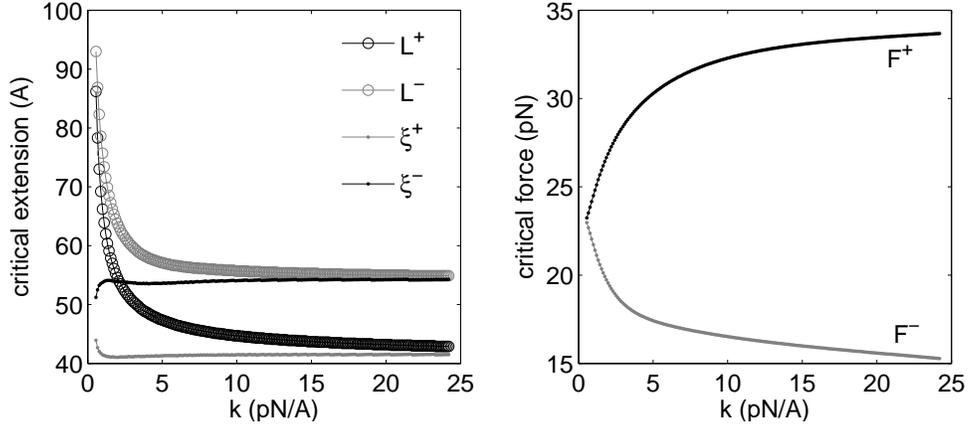}
\caption{Dependence of the instability of the $F$-$L$ isotherms  on the cantilever spring constant $k$ for the [3]rotaxane. The right panel shows the critical values of the force as a function of $k$. Here $F^+$ and $F^-$ correspond to the values of the force when the $F$-$L$ curves exhibit a maximum and minimum,  respectively. The left panel shows the  extension lengths  $L^\pm$  that enclose the unstable region in the $F$-$L$ isotherms,  as well as the  average end-to-end distance  at the critical points $\xi^\pm$. } 
     \label{fig:criticalforcevsk}
\end{figure}

Note that the  instability properties of  the $F$-$L$ isotherms  depend intricately on the cantilever spring constant employed. This dependence is succinctly conveyed in \fig{fig:criticalforcevsk} which shows the values of the extension  and the average force  at the critical points in the $F$-$L$ isotherms. In the figure,  $F^+$ and $L^+$ (or $F^-$ and $L^-$) denote the values of the force and the cantilever extension when the $F$-$L$ curve exhibits a maximum (or minimum). These values  enclose the unstable region in the $F$-$L$ isotherm. In turn, $\xi^{\pm}$ is the average end-to-end distance at the critical points which is approximately independent of $k$ except for very small $k$.   The critical forces $F^+$ and $F^-$  show a  smooth and strong dependence on the cantilever spring constant. For large $k$  the system persistently  shows a  region of instability in the isotherms. However, as the cantilever spring is made softer, $F^+$ and $F^-$  approach each other and for small $k$ the mechanical instability in the $F$-$L$ isotherms is no longer present. 

\subsubsection{Disappearance of the mechanical instability in the soft spring limit}

The disappearance of the instability regions in the $F$-$L$ measurements when the system is pulled with a soft cantilever is a general feature of the extension of any molecular system provided that no covalent bond breaking occurs during the process.  To see this, consider the soft-spring approximation of the configurational partition function of the system plus cantilever [\eq{eq:partfuncL2}]:
\be
\label{eq:softspringZ}
\frac{Z(L)}{Z_0} \approx \frac{\exp( -\beta k L^2/2 )}{Z_0} \int \ud \xi \exp[ -\beta\phi(\xi) ] \exp( \beta k L \xi ) =  \exp( -\beta k L^2/2 )\langle  \exp( \beta k L \xi ) \rangle,
\ee
where the notation $\langle f \rangle$ stands for the unbiased (cantilever-free) average of $f$. In writing \eq{eq:softspringZ} we have supposed that in the region of relevant $\xi$ (in which the integrand is non-negligible)  $k\xi^2/V_L(\xi)\ll 1$ and, hence, that the cantilever potential is well approximated by $V_L(\xi) \approx \frac{k L^2}{2} - kL \xi$. This approximation is valid provided that that no bond breaking is induced during pulling and permits the introduction of a cumulant expansion~\cite{cumulant} in the configurational partition function. Specifically, the average $\langle  \exp( \beta k L \xi ) \rangle$ can be expressed as:
\be
\label{eq:cumulants}
 \langle \exp( \beta k L \xi ) \rangle = \sum_{n=0}^{\infty} \frac{(\beta k L)^n}{n!} \langle \xi^n\rangle  =  \exp \left\{ \sum_{n=1}^{\infty}  \frac{(\beta k L)^n}{n!}  \kappa_n (\xi) \right\} 
\ee
where $\kappa_n(\xi)$ is the $n$-th order cumulant. In view of  Eqs.~\eqref{eq:averageforce}, \eqref{eq:softspringZ} and  \eqref{eq:cumulants} the slope of the $F$-$L$ curves can be expressed as:
\be
\frac{\partial \langle F\rangle_L}{\partial L} = k - \beta k^2 \sum_{n=0}^{\infty} \frac{(\beta k L)^n}{n!} \kappa_{n+2}(\xi).
\ee
It then follows that to lowest order in $k$,
\be
\frac{\partial \langle F \rangle_L}{\partial L} \approx k > 0.
\ee
That is, no unstable region in the $F$-$L$ curve can arise in the soft-spring limit, irrespective of the specific form of $\phi(\xi)$. Instabilities, however,  can arise in higher orders in the expansion. For instance, to second order in $k$ instabilities arise provided that the cantilever spring constant satisfies:
\be
k > \frac{1}{\beta \kappa_2(\xi)},
\ee 
where $\kappa_2(\xi) = \langle \xi^2 \rangle  - \langle \xi \rangle^2$.

\subsubsection{Emergence of the mechanical instability in the stiff spring limit}

A mechanically unstable region in the $F$-$L$ curves can only develop if the PMF of the unbiased molecule  by itself has a region of concavity. To understand how  this result arises consider  the configurational partition function of the molecule plus cantilever at extension $L$ [\eq{eq:partfuncL2}] for large $k$. In this regime,  most contributions to the integral will come from the region where $\xi = L$. Consequently,   $\exp [ -\beta \phi(\xi) ]$ can be expanded around this point  to give:
 \be
\label{eq:stiffexpansion}
 \exp[-\beta \phi(\xi)]  = \exp [-\beta \phi(L) ] \left[  1 - \beta A_1(L) (\xi - L) 
 -\frac{\beta}{2} A_2(L)  (\xi -L)^2  + \cdots\right],
 \ee
 where
 \be
 \begin{split}
A_1(L) & =  \frac{\partial \phi(L)}{\partial L} ,  \\
A_2(L) & =   \frac{\partial^2 \phi(L)}{\partial L^2} - \beta  \left( \frac{\partial \phi(L)}{\partial L} \right)^2.
 \end{split}
 \ee
Here,  the notation is such that  
$ \frac{\partial \phi(L)}{\partial L} = \left.  \frac{\partial \phi(\xi)}{\partial \xi}\right|_{\xi = L}$. Introducing \eq{eq:stiffexpansion}  into \eq{eq:partfuncL2},  integrating explicitly the different terms,  and performing an expansion around $1/k = 0$ one obtains:
 \be
 Z(L) = \sqrt{\frac{2\pi}{k\beta}} \exp[ -\beta \phi(L) ]\left[ 1   -\frac{1}{2 k }  A_2(L)     + \mc{O}(1/k^2) \right].
 \ee
Using \eq{eq:averageforce}, the derivative of the force can be obtained from this approximation to the partition function. To lowest order in  $1/k$ it is given by:
 \be
\frac{\partial \langle F \rangle_L}{\partial L } = \frac{\partial^2 \phi(L)}{\partial L^2} + \frac{1}{2\beta k} \frac{\partial^2 A_2(L)}{\partial L^2} + \mc{O}(1/k^2), 
 \ee
from which it follows that, to zeroth order in $1/k$, 
 \be
 \label{eq:stiffspring}
\frac{\partial \langle  F \rangle_L}{\partial L } = \frac{\partial^2 \phi (L)}{\partial L^2} + \mc{O}(1/k).
\ee
That is, an unstable region in the  force vs. extension requires a region of concavity in the PMF.

\subsubsection{Relation between force fluctuations and mechanical stability}
\label{stn:flucsandinstability}
The mechanical stability properties observed during the AFM pulling experiments are intimately   related to the fluctuations in the force measurements. To see this,  note that  the  average slope of the $F$-$L$ curves can be expressed in terms of the  fluctuations in the force:
 \be
\label{eq:ffluctuations}
 \frac{\partial \langle F \rangle_L}{\partial L} = \left  \langle \frac{\partial^2 V_L(\xi)}{\partial L^2} \right \rangle_L
      - \beta [\langle F^2 \rangle_L - \langle F \rangle_L^2] =  k 
      - \beta [\langle F^2 \rangle_L - \langle F \rangle_L^2],
 \ee
where we have employed \eq{eq:averageforce}.
The sign of $ \frac{\partial \langle F \rangle_L}{\partial L}$   determines the mechanical stability during the extension and hence \eq{eq:ffluctuations}  relates the thermal fluctuations in the force measurements with the  stability properties of the $F$-$L$ curves. Specifically, for the stable branches for which $ \frac{\partial \langle F \rangle_L}{\partial L} >0$ the force fluctuations satisfy
\begin{subequations}
\label{eq:inequalities}
 \be
 \langle F^2 \rangle_L - \langle F \rangle_L^2 < \frac{k}{\beta}.
 \ee
In turn, at the critical points where the derivative changes sign the force fluctuations satisfy a strong constraint:
\be
 \langle F^2 \rangle_L - \langle F \rangle_L^2 = \frac{k}{\beta}.
 \ee
Last, in  the unstable branches the force fluctuations are larger than in the stable branches and satisfy the inequality
\be
\label{eq:unstable}
 \langle F^2 \rangle_L - \langle F \rangle_L^2 > \frac{k}{\beta}.
\ee
\end{subequations}
 These inequalities can be employed to determine critical points in the force by measuring the fluctuations even in situations where the fluctuations mask the presence of critical points.   

\begin{figure}[htbp]
\centering
\includegraphics[width=0.8\textwidth]{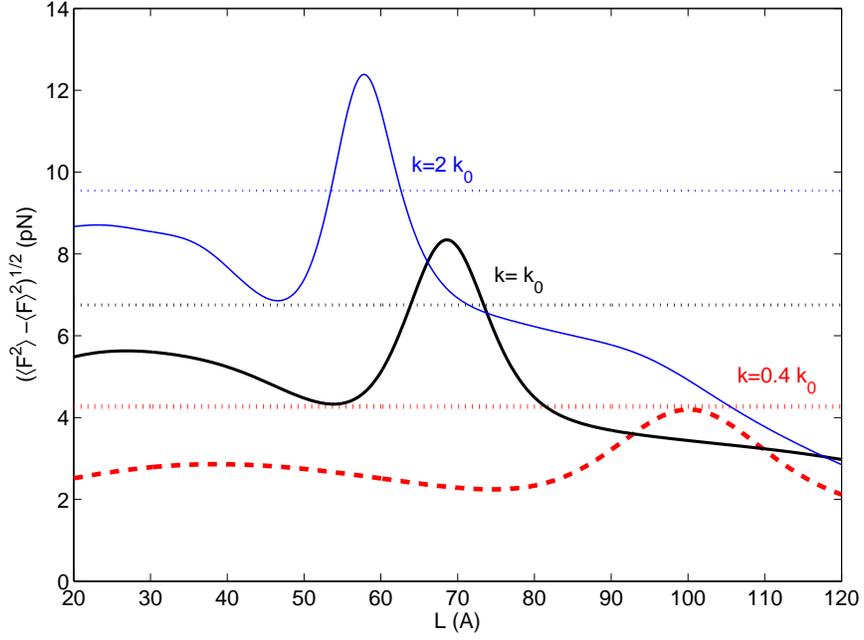}
\caption{Standard deviation  in the force measurements $\sigma_F = \sqrt{\langle F^2 \rangle_L - \langle F \rangle_L^2}$ as a function of $L$ for three different cantilever spring constants $k$ during the pulling of the [3]rotaxane. In each case, the dotted line indicates the value of $\sqrt{k/\beta}$ which sets the limit between the stable and unstable branches  in the extension, see \eq{eq:inequalities}. }
     \label{fig:fluctuations}
\end{figure}

Figure~\ref{fig:fluctuations} illustrates these general observations in  the specific case of the pulling of [3]rotaxane. As shown, for  $k=k_0$ and $k=2k_0$ there is a region where the force fluctuations become larger than $k/\beta$ and, consequently, unstable behavior in $F$-$L$ develops. For  $k=0.4k_0$ or less the fluctuations in the force are never large enough to  satisfy \eq{eq:unstable} and no critical points in the $F$-$L$ develop, as can be confirmed in \fig{fig:criticalforcevsk}.

\subsubsection{Origin of the blinking in the force measurements}
\label{stn:dynamicalbistability}

\begin{figure}[htbp]
\centering
\includegraphics[width=0.8\textwidth]{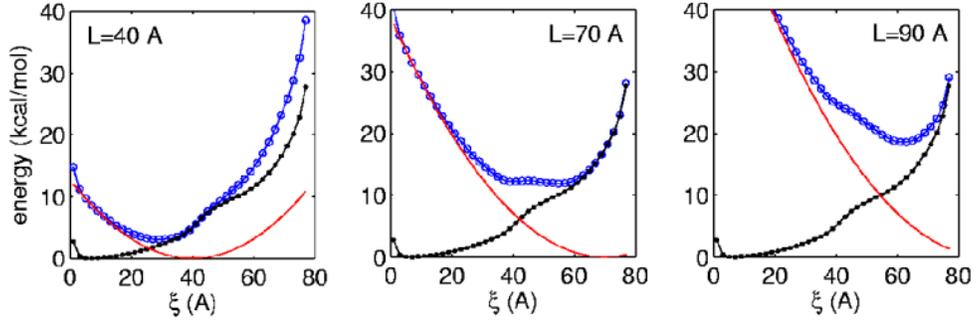}
\caption{Effective potential $U_L(\xi)=\phi(\xi) + V_L(\xi)$  for the molecule plus cantilever  for different values of the extension $L$. In the panels, the open circles correspond to $U_L(\xi)$, the full circles to the PMF $\phi(\xi)$, and the solid lines to the cantilever potential  $V_L(\xi)$.   The cantilever spring constant  employed is the same as the one used in the pulling simulations presented in \stn{stn:phenomenology}. Note the bistability in the effective potential for $L=70$ \AA.  } 
     \label{fig:emergencebistability}
\end{figure}

The observation of  force measurements that blink between a high-force and a low force regime (recall \fig{fig:3DNP2ring.unstable}) requires the composite system to be bistable along the end-to-end distance for some $L$, i.e. the effective potential $U_L(\xi)$ [\eq{eq:effectivepotential}] must have a double minimum.  Since the PMF of the molecule by itself is not bistable along $\xi$ (see \fig{fig:3DNP2ring.molecularphi}), the bistability must be introduced by the cantilever potential. 
 Figure~\ref{fig:emergencebistability} shows $U_L(\xi)$ for selected $L$ for the [3]rotaxane. For $L$ in the stable regions of the free energy ($L=40$ \AA~ and $L=90$ \AA) the  effective potential exhibits a single minimum along the $\xi$ coordinate.  However, for $L=70$ \AA ~a bistability in the potential develops. The secondary minimum  is the cause  for the blinking in the force measurements observed at this $L$ (cf. \fig{fig:3DNP2ring.unstable}). Further, the relative energy between the two induced minima around the transition region can be manipulated by varying $L$.

What are the minimum requirements for the emergence of bistability along $\xi$? A necessary condition  is that the effective potential $U_L(\xi)$ is concave for some region along $\xi$, that is:
\be
\label{eq:bistability}
\frac{\partial^2 U_L(\xi)}{\partial \xi^2} = \frac{\partial^2 \phi(\xi)}{\partial \xi^2} + k < 0.
 \ee
For \eq{eq:bistability} to be satisfied it is required that both: (i) the  PMF of the isolated molecule has  a region of concavity where $\frac{\partial^2 \phi(\xi)}{\partial \xi^2} < 0$, and (ii) the cantilever employed is sufficiently soft such that 
\be
\label{eq:bistabilityk}
k < -\ti{min} \left(\frac{\partial^2 \phi(\xi)}{\partial \xi^2}\right),
\ee
for some $\xi$.  Equation~\eqref{eq:bistabilityk} imposes an upper bound on $k>0$ for bistability to be observable. If $k$ is stiff the  inequality would be violated for all $\xi$ and bistability would not be manifest. Note, however,  that there is no lower bound for $k$ that prevents bistability along $\xi$. This is in stark contrast with the behavior of the  mechanical instabilities which disappear  in the soft-spring limit and persist when stiff springs are employed (see Secs.~\ref{stn:effectofk}-\ref{stn:flucsandinstability}).

\begin{figure}[htbp]
\centering
\includegraphics[width=1.0\textwidth]{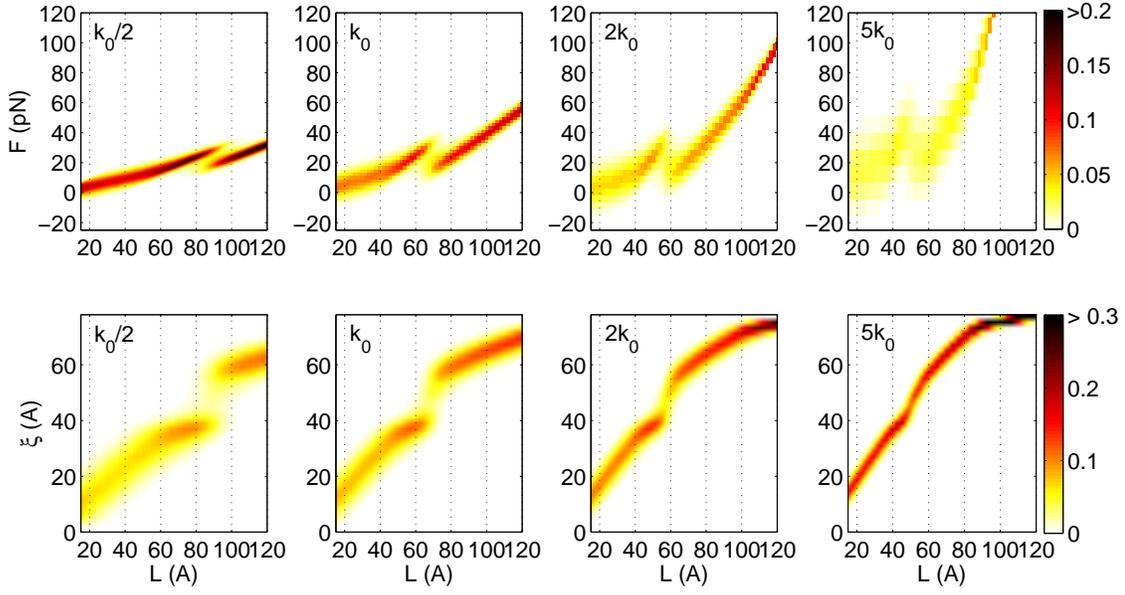}
\caption{The upper panels show the probability density distribution of the force measurements $p_L(F)= \langle \delta[F - F(\vect{r}, L)]\rangle_L$ (in pN$^{-1}$)  during the extension of  the [3]rotaxane using cantilevers of varying stiffness.  The force function $F(\vect{r},L)$ is defined by \eq{eq:force}. The lower panels show the associated spatial probability density distributions $p_L(\xi)$ (in \AA$^{-1}$) [\eq{eq:biased}] for the [3]rotaxane plus cantilever.   The color code is given in the far right.  The spring constants are expressed in terms of $k_0=1.1$ pN/\AA, the value employed in the pulling simulations presented in \stn{stn:phenomenology}.    Note how bistability along  $\xi$, and hence blinks in the force measurements, arises for soft cantilevers and decays for stiff ones in accordance with  \eq{eq:bistabilityk}  } 
     \label{fig:bistabilityall}
\end{figure}

 Figure~\ref{fig:bistabilityall}  illustrates these general observations during the pulling of  the [3]rotaxane.  Specifically,  the figure shows, for cantilevers of varying stiffness, the probability density of observing the value $F$ in the force measurements $p_L(F)$ (upper panels) and the associated probability density $p_L(\xi)$ that the molecule adopts the extension $\xi$    (lower panels) when the cantilever is fixed at $L$. For soft springs the system  satisfies inequality \eqref{eq:bistabilityk} for a range of $L$ values  and there is a clear bistability in both probability distributions.  As $k$ is increased this bistability becomes less prominent and  for $k=5k_0$  is no longer observed.

\subsubsection{Final remarks}

\begin{figure}[htbp]
\centering
\includegraphics[scale=0.4]{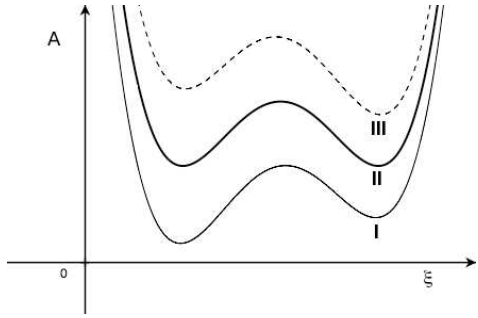}
\caption{Schematic variation of the Helmholtz potential of the molecule plus cantilever as a function of $\xi$  for different extensions $L$ suggested by the phenomenological observations around the region of mechanical instability.  The labels I-III correspond to the different stability regions in \fig{fig:3DNP2ring.freeenergysc}}. 
     \label{fig:phasetransition}
\end{figure}

What  then should be the picture of the pulling? Clearly, a picture like the one shown in \fig{fig:phasetransition}, although appropriate to describe  the phenomenology observed in \stn{stn:phenomenology} around the transition region (and the basis of an influential model of molecular extensibility~\cite{reif}),  is  not accurate since  the molecule itself does not show a double well structure along $\xi$ in the free energy. Note, however, that the molecular free energy is inherently bistable along some natural unfolding pathway. If one were able to determine such a pathway,  a picture reminiscent to the one shown in \fig{fig:phasetransition} would emerge. However, such natural unfolding pathway is not necessarily accessible  during the pulling experiment.

The analogy between first-order phase transition and the isometric single-molecule pulling experiments, even when suggestive, is limited by the fact that the qualitative features observed during pulling strongly depend on the cantilever constant employed. This is because the properties that are measured are those of the combined cantilever plus molecule system.
However, as previously noted~\cite{kreuzer2001},  in the stiff spring limit the measured $F$-$L$ curves resemble those which  would be obtained  from a calculation in the Helmholtz ensemble of the isolated molecule. In this limit, the instability behavior in the force vs. extension will only be due to molecular properties  [recall \eq{eq:stiffspring}], and the associated bistability is that observed along the radius of gyration when $\xi$ is fixed in the concave region of the PMF (\fig{fig:unstablenocantilever}). It is in this limit that the phenomenology observed during the isometric single-molecule pulling is closest to a first-order phase transition. Experimentally, this limit is impractical since employing a stiff spring reduces the sensitivity in the force measurements.

\section{Conclusions}
\label{stn:conclusions}

In this contribution we have simulated the equilibrium isometric AFM pulling of a series of donor-acceptor oligorotaxanes by means of constant temperature molecular dynamics in a high dielectric medium using MM3 as the force field. For this system, hysteresis and other nonequilibrium effects that may arise during  pulling can be overcome by reducing the pulling speed.  The resulting equilibrium force vs. extension isotherms show a mechanically unstable region in which the molecule unfolds and, for selected extensions, blinks in the force measurements between a high-force and a low-force regime.  

From the force vs. extension data,  the Helmholtz free energy profile  for the oligorotaxanes along the extension coordinate was reconstructed using the weighted histogram analysis method, and decomposed into energetic and entropic contributions. The simulations reveal an unfolding pathway for the oligorotaxanes, and  indicate that the folding  is energetically driven but entropically penalized.   Even when the energy and entropy contributions to the potential of mean force can vary widely along the extension, these effects largely cancel one another leading to only modest changes in the free energy profile. Further, we have observed that, for the rotaxanes studied, increasing the number of threaded rings stabilizes the folded conformation, making it resilient to unfolding over a wider range of end-to-end distances. As expected, the elastic properties of the extended conformation are not affected by the number of threaded rings.  

\begin{figure}[htbp]
\centering
\includegraphics[width=0.8\textwidth]{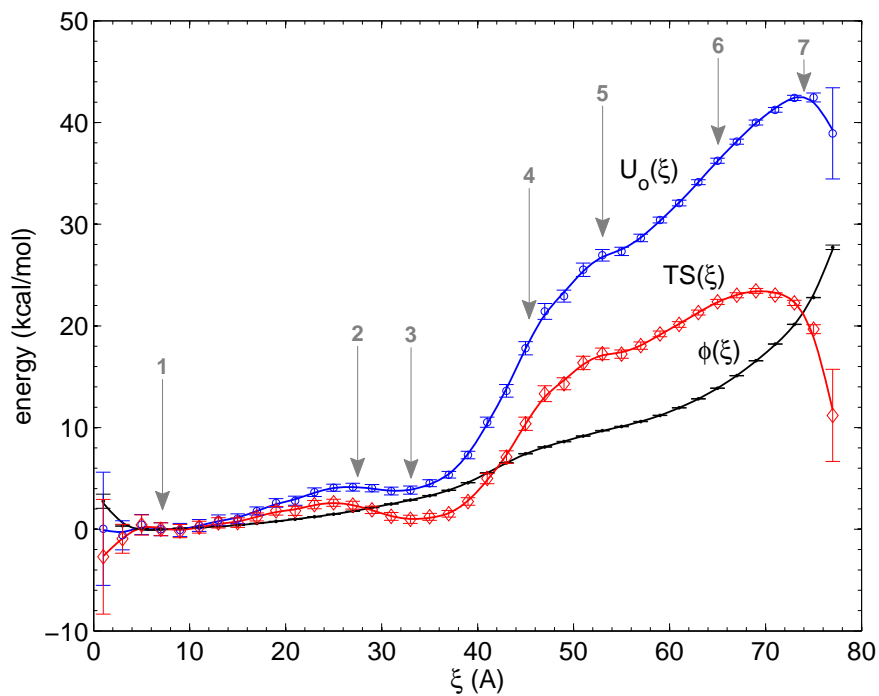} 
\caption{Potential $U_0(\xi)$ and entropic $TS(\xi)$ contributions to the PMF [\eq{eq:decomposition}] of the [3]rotaxane along the end-to-end distance $\xi$. The results are averages of three different pulling simulations.  The error bars correspond to twice the standard deviation obtained from a bootstrapping analysis. The solid lines result  from a spline interpolation of the available data points. Typical structures (labels 1-7) are shown in \fig{fig:3DNP2ringsnapshots}.  }
     \label{fig:3DNP2ring.molecularphi}
\end{figure}

\begin{figure}[htbp]
\centering
\includegraphics[width=0.8\textwidth]{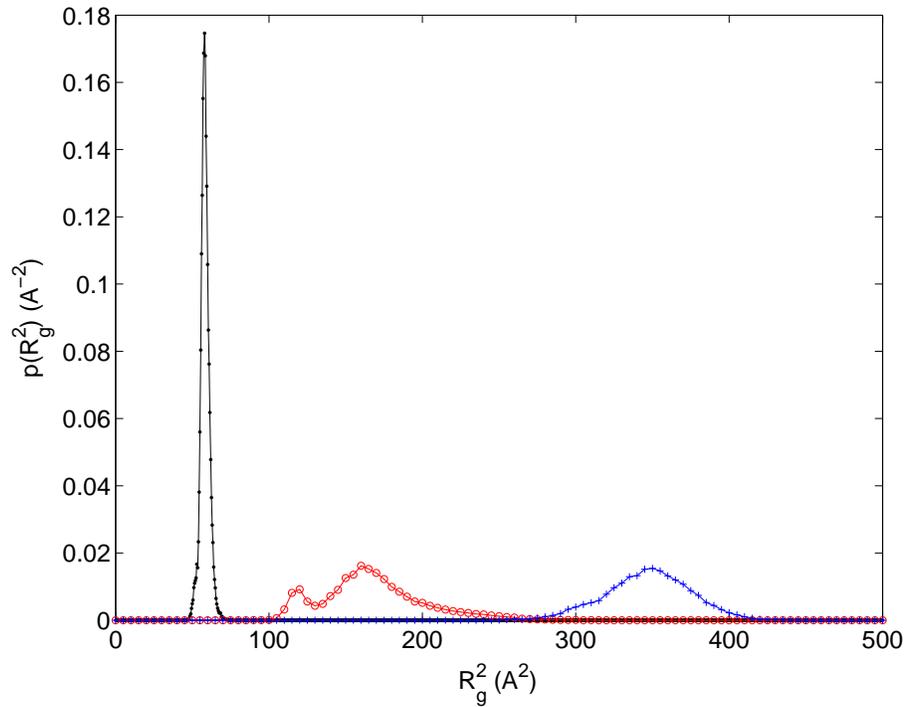}
\caption{Probability density distribution of the radius of gyration $p(R_g^2)$ for the [3]rotaxane when $\xi$ is fixed at 7.0 \AA~(dots), 47~\AA~(open circles) and 64.9 \AA~(crosses).  Note the bistability along $R_g^2$ when $\xi$ is fixed at the concave region of the PMF. }. 
     \label{fig:unstablenocantilever}
\end{figure}

The potential of mean force of the rotaxanes  exhibits only one minimum along the end-to-end distance. It consists of two stable convex regions characterizing the folded and unfolded conformations and a region of concavity where the unfolding occurs. 
This thermodynamically unstable region  arises because the molecule is inherently bistable. The inherent molecular bistability is not resolved  along the end-to-end coordinate  (cf. \fig{fig:3DNP2ring.molecularphi}) but, at least in the cases considered, it is manifest along the radius of gyration  (recall \fig{fig:unstablenocantilever}). 

Clearly,  modifying the nature of  the solvent employed during the pulling can have a quantitative effect on the determined free energy profile since the solvent can vary  the relative stability of the different molecular conformations encountered along the extension. Nevertheless, except for very good solvents, the qualitative features of the topography of the  potential of mean force described above are expected to remain intact. Also, hydrodynamic effects  that may appear when considering an explicit solvent  (e.g. drag forces acting on the AFM cantilever due to viscous friction with the surrounding solvent~\cite{janovjak})  are immaterial for the equilibrium pulling.

In addition, general conditions on the molecule and on the cantilever stiffness required for the emergence of mechanical instability and blinking in the force measurements in isometric pulling experiments have been presented. Specifically, we have shown that for such effects to arise the potential of mean force of the molecule needs to exhibit a region of concavity along the end-to-end distance. Further, the stiffness of the cantilever employed needs to satisfy specific constraints.  For the blinks in the force measurements to arise the cantilever spring  constant needs to be sufficiently soft so that   \eq{eq:bistabilityk} is satisfied. In turn,  for the mechanical instability to be observable the fluctuations in the force measurements have to be sufficiently large to satisfy  \eq{eq:unstable}  for some $L$. In practice, this implies that in the stiff-spring limit the mechanical instability can emerge, while in the soft spring limit the mechanical instability decays regardless of the details of the molecular system, provided that no covalent bond-breaking occurs during  pulling. 

In this light, it becomes apparent that the analogy presented between first-order phase transitions and the mechanical properties observed during the extension is limited by the fact that the observed behavior depends on the cantilever constant employed. The analogy is the closest in the stiff-spring limit where the mechanical properties recorded are those  that would have been obtained in the Helmholtz ensemble of the isolated molecule.

\begin{acknowledgments}
The authors thank Prof. J. Fraser Stoddart,  Dr. Wally F. Paxton and Dr. Subhadeep Basu for  insightful remarks, and  the AFOSR-MURI program of the DOD,  NSF Grant CHE-083832 and the 
Network for Computational Nanoscience for support of this work. I.F. thanks Dr. Gemma C. Solomon and Dr. Neil Snider for discussions on an earlier version of this manuscript.
\end{acknowledgments}


\clearpage

\clearpage

\clearpage

\clearpage

\clearpage


\clearpage

\clearpage

\clearpage

\clearpage


\clearpage

\clearpage


\clearpage

\clearpage

\clearpage


\clearpage

\clearpage

\clearpage

\end{document}